\documentclass[aps,twocolumn,prd,epsf,showpacs,amsmath,amssymb]{revtex4}
\usepackage{dcolumn}
\usepackage{bm}
\newcommand{\e}{equation$\;$}

\newcommand{\be}{\begin{equation}}
\newcommand{\ee}{\end{equation}}
\newcommand{\ba}{\begin{eqnarray}}
\newcommand{\ea}{\end{eqnarray}}
\newcommand{\ban}{\begin{eqnarray*}}
\newcommand{\ean}{\end{eqnarray*}}      
\newcommand{\n}[1]{\label{#1}}

\newcommand{\eq}[1]{(\ref{#1})}

\newcommand{\ntr}{\ensuremath{{\nu(t,r)}}}
\newcommand{\str}{\ensuremath{\psi(t,r)}}
\newcommand{\ph}{\ensuremath{{\phi}}}
\newcommand{\dw}{\ensuremath{{d\Omega^2}}}

\newcommand{\E}{{\mathbb E}}

\begin{document}
\title{Trapped surfaces and horizons in static massless scalar field spacetimes }
\author{Swastik Bhattacharya$^1$ and Pankaj S. Joshi$^1$}
\email{swastik@tifr.res.in, psj@tifr.res.in}
\affiliation{$^1$Tata Institute for Fundamental Research, Homi Bhabha Road,
Mumbai 400005, India}

\begin{abstract}
We consider here the existence and structure of trapped surfaces, 
horizons and singularities in spherically symmetric static massless scalar 
field spacetimes. Earlier studies have shown that there exists no event horizon 
in such spacetimes if the scalar field is asymptotically flat.  We extend 
this result here to show that this is true in general for spherically 
symmetric static massless scalar field spacetimes, whether the scalar field 
is asymptotically flat or not. Other general properties and certain 
important features of these models are also discussed.
\end{abstract}


\maketitle

\section{Introduction}
Spherically symmetric solutions of Einstein equations for
static massless scalar field configurations have been investigated in 
considerable detail in the past. There are two different ways in which this 
has been done. The first approach has been to look for solutions of 
the Einstein equations in this case and thus gain more understanding
on the structure of the solutions. Bergmann and Leipnik 
\cite{Bergman} 
were among the first to construct such spherically symmetric static 
solutions for massless scalar fields. They had, however, only
a limited success due to an inappropriate choice of coordinates. 
Around the same time Buchdal 
\cite{Buchdal} 
developed techniques to generate solutions for this system, and also  
Yilmaz 
\cite{Yilmaz} 
and Szekeres 
\cite{Szekeres} 
found some classes of solutions for the static massless scalar 
field configurations in general relativity. Subsequently, Wyman 
\cite{Wyman} 
systematically discussed these solutions and showed 
a general method to obtain solutions in the case when the 
scalar field was allowed to have no time dependence. This gave a 
unified method to obtain most of the solutions obtained earlier.
Also, Xanthopoulos and Zannias 
\cite{Xan} 
gave a class of solutions for time independent scalar fields 
in arbitrary dimensions
where the spacetime metric was static.

Further, the scalar fields conformally coupled 
to gravity have been a subject of immense interest to many 
researchers 
\cite{conformal}.
Static massless scalar fields have also been investigated
in settings more general as compared to spherical symmetry
(see e.g.
\cite{scal}).
Thus, there have been many investigations on special cases 
of static solutions of the Einstein equations for the massless 
scalar field system, as indicated above.

It is clear, however, from the analysis of Wyman 
\cite{Wyman}, 
that there is still a large class of solutions that 
as yet remains unexplored. In fact, Wyman identifies 
that class to be the solutions of a particular non-linear ordinary
differential equation, which is difficult to solve analytically. 
It turns out nevertheless that even in these cases when it is not possible 
to solve the Einstein equations explicitly or completely, there are still 
some general properties of such solutions that can be deduced 
which are of physical interest. This is the second way of approaching 
this problem.  In that direction, some interesting general properties 
of static massless scalar field spacetimes have been found 
by Chase 
\cite{Chase}. 
For example, it was found that every massless scalar 
field, which is gravitationally coupled and asymptotically flat, becomes 
singular at a simply-connected event horizon. In fact, the result 
obtained by Chase is more general in the sense that it does 
not assume spherical symmetry of the spacetime. However, for 
this result to hold true, the scalar field has to be asymptotically flat
necessarily, which means that the scalar field goes to zero in the limit 
of going to spatial infinity. But there can be solutions, where this 
condition does not apply. In fact, we shall show later that such solutions 
exist for static spherically symmetric massless scalar 
field spacetimes.

One of the main interest for the study of these solutions 
comes from the fact that for simple models for gravitational collapse 
in general relativity, these  settle to a Schwarzschild solution 
where the final singularity of collapse is hidden within an event horizon 
of gravity. But the analogous static scalar field solutions do not 
have such event horizons, as was pointed out by many of the above 
works. Of course, such models cannot be considered to be 
counter-example to the cosmic censorship conjecture, which states 
that naked singularities do not develop as final state of 
gravitational collapse. That is because these are static solutions 
and do not develop from collapse from a regular initial data. 
However, the cosmic censorship hypothesis continues 
to be unproved and without even a definite mathematical formulation, 
despite serious attempts in that direction for past many decades.
Therefore a study of models such as undertaken here would be 
of interest as this may provide us with an understanding into 
the nature and structure of the singularities, horizons, and trapped 
surfaces that can take place or exist in general relativistic 
spacetime models.

From such a perspective,  in this note we extend 
the result found by Chase in the case of static spherically symmetric 
massless scalar field spacetimes, by showing that in such 
spacetimes, event horizon cannot exist, whether the scalar field is 
asymptotically flat or not. However, our result holds true and 
extends the findings by Chase only for spherically symmetric 
spacetimes, whereas Chase had not assumed  any such symmetry 
of the spacetime, apart from its staticity. Apart from this  
result, we also discuss briefly the existence of singularities 
for these models and their visibility.

In section II, we write down the Einstein 
equations for the massless scalar fields in comoving 
coordinates. As explained there, the spacetimes in this case are 
divided into two classes, namely, $\phi=\phi(t)$ and $\phi=\phi(r)$. 
This coordinate system has been used to analyse the static 
massless scalar field models by Wyman
\cite{Wyman}
and also others. In section III, we shall consider 
the general properties of both these classes of models. 
Specifically, we shall show that if the spacetime is non-empty, 
then event horizon does not exist. In the final Section we 
give and summarize the conclusions.

\section{The Einstein equations}

In our analysis here, we consider a four-dimensional spacetime 
manifold which has spherical symmetry.
The massless scalar field $\ph(x^a)$ on such a spacetime 
$(M, g_{ab})$ is described by the Lagrangian, 
\be
{\cal L}=-\frac{1}{2}\ph_{;a}\ph_{;b}g^{ab}.
\n{lag}
\ee
The corresponding Euler-Lagrange equation is then given by, 
\be
\ph_{;ab}g^{ab}=0,
\n{el}
\ee
and the energy-momentum tensor for the scalar field, as calculated 
from the above Lagrangian, is given as 
\be
T_{ab}=\ph_{;a}\ph_{;b}-\frac{1}{2}g_{ab}\left(\ph_{;c}\ph_{;d}
g^{cd}\right).
\n{emt}
\ee
The massless scalar field is a {\it Type I} matter field
\cite{haw},  
{\it i.e.}, it admits one timelike and three spacelike 
eigen vectors. At each point $q\in M$, we can express the tensor 
$T^{ab}$ in terms of an orthonormal basis $(\E_0,\E_1,\E_2,\E_3)$, 
where $\E_0$ is a timelike eigenvector with the eigenvalue $\rho$,
and $\E_{\alpha}$ $(\alpha=1,2,3)$ are three spacelike eigenvectors 
with eigenvalues $p_\alpha$. The eigenvalue $\rho$ represents 
the energy density of the scalar field as measured by an observer 
whose world line at $q$ has an unit tangent vector $\E_0$, and 
the eigenvalues $p_\alpha$ represent the principal pressures 
in three spacelike directions $\E_\alpha$.

We now choose the spherically symmetric coordinates 
$(t,r,\theta,\phi)$ along the eigenvectors $(\E_0,\E_\alpha)$, 
such that the reference frame is {\it comoving}.  This coordinate 
system has also been used by Wyman 
\cite{Wyman} 
to discuss the 
static massless scalar field spacetimes. As discussed in 
\cite{landau}, 
the general spherically symmetric metric in comoving 
coordinates can be written as,
\begin{equation}
ds^2= e^{2\ntr}dt^2-e^{2\str}dr^2-R^2(t,r)\dw,
\label{metric}
\end{equation}
where $\dw$ is the metric on a unit two-sphere and 
we have used the two gauge freedoms of two variables, namely, 
$t'=f(t,r)$ and $r'=g(t,r)$, to make the $g_{tr}$ term in the 
metric and the radial velocity of the matter field to vanish. 
That means that the energy-momentum tensor is necessarily 
diagonal in such a coordinate system.
We note that we still have two scaling freedoms of one variable
available, namely $t \to f(t)$ and $r \to g(r)$. We note
here that the metric function $R$ is sometimes called the 
physical radius, especially in gravitational collapse 
situations.

As we are considering here spherically symmetric spacetimes, 
we have $\ph=\ph(t,r)$ necessarily.
Furthermore, from  \e\eq{emt} we can easily see that in the 
comoving reference frame with the metric given by \eq{metric},
$T_{10}=\phi' \dot{\phi}$. It follows therefore that
we must have here necessarily 
$\ph(t,r)=\ph(t)$ or $\phi(t,r)=\phi(r)$, 
where the energy-momentum tensor is necessarily diagonal.

For the metric (\ref{metric}),
and using the following definitions,
\begin{equation}
G(t,r)=e^{-2\psi}(R^{\prime})^{2},\;\; H(t,r)=e^{-2\nu} (\dot{R})^{2}\;,
\n{eq:ein5}
\end{equation}
\begin{equation}
F=R(1-G+H)\;,
\label{eq:ein4}
\end{equation}
we can write the independent Einstein equations for the 
spherical massless scalar field (in the units $8\pi G=c=1$) 
as below (see \cite{GJ1}),
\begin{equation}
 \rho=\frac{F'}{R^2 R'},
\end{equation}
\begin{equation}
 P_r=-\frac{\dot{F}}{R^2 \dot{R}},
\end{equation}
\begin{equation}
 \nu'(\rho+P_r)=2(P_\theta-P_r)\frac{R'}{R}-P_r'\label{e3},
\end{equation}
\begin{equation}
 -2\dot{R}'+R'\frac{\dot{G}}{G}+\dot{R} \frac{H'}{H}=0\label{e1}, 
\end{equation}

In the above, the function $F(t,r)$ has the interpretation 
of the mass function for the matter field, in that it represents 
the total mass contained
within the sphere of coordinate radius $r$ at any given time $t$. 
As noted above, in the static case the metric components 
$g_{\mu\nu}$s are functions of the radial coordinate $r$ only necessarily, 
but the scalar field $\phi$ itself can still be either 
$r$ or $t$ dependent. In either of these cases, namely
$\phi=\phi(r)$ or $\phi=\phi(t)$, the equation 
of state relating the energy density and pressure for the 
scalar field are different as we shall find here. 
We note that the condition of staticity for
the spacetime metric implies that the metric components are not 
functions of time. There is no such restriction, however, on 
the scalar field itself, which can still be time-dependent.

\section{General properties of the spacetimes}

We shall now analyse the Einstein equations given above for
the massless scalar field, and point out several properties of 
these solutions in general.  Many of these relate to the nature of 
the singularity and the trapped surfaces in these models.
The Einstein equations are not fully solved here but a general
analysis of their properties is carried out which implies these 
conclusions, which thus hold true for all solutions of this
system.

\subsection{The $\phi=\phi(r)$ class of models}

In this case for static spacetimes, when $\phi=\phi(r)$ 
and $g_{\mu\nu}= g_{\mu\nu}(r)$, 
the Einstein equations given in the previous section reduce 
to the following set of equations,
\begin{equation}
\frac{1}{2}e^{-2 \psi} \phi'^2 = \frac{F'}{R^2 R'},
\end{equation}
\begin{equation}
\frac{1}{2}e^{-2\psi} \phi'^2 = e^{-2\psi}(\frac{R'^2}{R^2}+\frac{2R'\nu'}{R})
-\frac{1}{R^2},
\end{equation}
\begin{equation}
\phi'' = (\psi'-\frac{2R'}{R}-\nu')\phi'\label{s1}.
\end{equation}
\begin{equation}
e^{-2\psi} R'^2 = 1-\frac{F}{R} 
\end{equation}
In the above, the equation \eqref{s1} can be integrated once 
with respect to $r$ to give
\begin{equation}
\phi'=\frac{e^{\psi-\nu+a}}{R^2}\label{fld},
\end{equation}
where $a=const.$

Eliminating now $\phi'$ from these equations gives,
\begin{equation}
\frac{1}{2}\frac{e^{-2\nu+2a}}{R^2} R' = F'\label{s2},
\end{equation}
\begin{equation}
\frac{1}{2}\frac{e^{-2\nu+2a}}{R^4} = e^{-2\psi}(\frac{R'^2}{R^2}+\frac{2R'\nu'}
{R})-\frac{1}{R^2} \label{s3}
\end{equation}
\begin{equation}
e^{-2\psi} R'^2 = 1-\frac{F}{R} \label{s4}
\end{equation}
We note that there is still a freedom left to transform the 
coordinate $R$, and thus the number of unknown variables is reduced 
to three in the three equations above. This freedom is just a 
coordinate transformation of the form 
\begin{equation}
 r\to g(r),
\end{equation}
which is allowed by the spherical symmetry of the spacetime.


The implications of the Einstein equations above will be 
investigated now. In what follows, we do not make any further 
assumptions or special choices and therefore the 
conclusions apply in generality. We define a function 
$f(R)$ by,
\begin{equation}
f(R),_R = \frac{e^{-2\nu}}{2R^2} \label{gic}
\end{equation}
Using this in \eqref{s2}, we get,
\begin{equation}
F=e^{2a}f(R)+C_1 
\end{equation}
We can take $C_1=0$, which gives $F=e^{2a}f(R)$. 
Using this in \eqref{s4}, we get
\begin{equation}
e^{-2\psi} R'^2 = 1-\frac{e^{2a}f(R)}{R}
\end{equation}
Also, from \eqref{gic} we get,
\begin{equation}
-2\nu'=(\frac{f,_{RR}}{f,_R}+\frac{2}{R})R'
\end{equation}

Taking $e^{2a}=1, C_1=0$ for simplicity and clarity 
of presentation, the last two equations, together 
with \eqref{s3} give
\begin{equation}
 R(f,_R)^2 = (f-R)(f,_R+Rf,_{RR})-Rf,_R \label{gi}
\end{equation}
The equation above holds true in generality and where
we have not made any special coordinate choices. 
This represents clearly the main Einstein equation in 
the case when $\phi=\phi(r)$, and the solutions to the same 
give the classes of allowed solutions in this case for the 
static massless scalar field in general relativity. 
The above is a non-linear
ordinary differential equation of second order 
which is in general difficult to solve fully. 
It is, however, possible to draw some general consequences 
from this, as we now show below. In particular, we point out 
the implications of the above towards the existence and nature
of the spacetime singularity and trapped surfaces 
in these spacetime models.

\subsubsection{No trapped surfaces in the spacetime}

For spherically symmetric spacetimes, the apparent horizon 
surface is given in general by the equation,
\begin{equation}
g^{\mu\nu} R,_\mu R,_\nu = 0 \label{aph}
\end{equation}
 We note here that for the static case, the apparent 
horizon and the event horizon are the same.
In the static case, the above becomes $e^{-2\psi}R'^2=0$. 
Using \eqref{s4}, this condition can be rewritten as $f(R)= R e^{-2a}$. 
When $ e^{2a} = 1$, we have
\begin{equation}
 f(R)=R .
\end{equation}
We note here that the results derived here do not 
depend upon changing the value of the constant $a$ and
their qualitative nature remains the same. Substituting the
above equation in 
\eqref{gi}, we get $Rf,_R(f,_R+1)=0$.
We note that because of the energy condition
ensuring the positivity of the mass energy density, namely 
$\rho \ge 0$, $f,_R$ cannot be negative. Further, in a 
non-empty spacetime, the density of the scalar field 
is non-zero everywhere. 
This implies that $f,_R \neq 0$ for $R>0$. 
It follows that the apparent horizon condition 
$f(R)=R$ can be satisfied only at $R=0$. 
Since the physical area is zero at $R=0$, in this 
case the event horizon or trapped surfaces
do not therefore exist in the spacetime.


\subsubsection{All solutions are singular at $R=0$}

In order to examine the existence of spacetime singularity 
in these models, we investigate the behaviour of the Ricci scalar, 
which is given here by the expression $R_c=- e^{-2\nu+2a}/R^4$. 
This gives,
\begin{equation}
 R_c=-\frac{2f,_Re^{2a}}{R^2} \label{Rscalar}
\end{equation}

We can see from the expression for the Ricci scalar, that 
if $R_c$ is to remain finite as $\lim{R \to 0}$, 
then clearly $\lim_{R \to 0}f,_R$ should go to zero atleast 
as fast as $R^2$. 
Therefore, let us consider the case when 
$\lim_{R \to 0}f(R)\sim R^\alpha$, where $\alpha \geq 3$. 
Then $f,_R\sim R^{\alpha-1}$ and $f,_{RR} \sim R^{\alpha-2}$. 
Keeping only largest terms as 
$\lim{R \to 0}$ in \eqref{gi}, 
we get $-R(f,_R+Rf,_{RR})-Rf,_R=0$ or
\begin{equation}
Rf,_{RR}+2f,_R=0
\end{equation}
From this, we get $f,_R \sim \frac{1}{R^2}$, which contradicts 
the initial assumption about $\lim_{R \to 0}f(R)$. 
This conclusion remains the same even if $f$ goes to zero faster 
than $R^2$, even in the cases when it is not necessarily a power 
law dependence on $R$. 
This means that \eqref{gi} cannot have any solutions such that 
$\lim_{R \to 0}f,_R$ goes to zero as $R^2$ or faster. This in turn 
implies that there will always be a spacetime singularity 
occurring at $R=0$,
for all the solutions in the static class of massless 
scalar field models with $\phi=\phi(r)$.


\subsubsection{Radial outgoing null geodesics from $R=0$}

In what follows, we show that there exist 
radial null geodesics coming out 
from the singularity at $R=0$, whenever the condition for apparent horizon 
is satisfied only at $R=0$ {\it i.e.} there is no apparent or event horizon 
in the spacetime, which is the case for all non-empty 
spacetimes as we pointed out above. 
The radial null geodesics in the spacetime are given
by the equation, $ds^2 = 0 = e^{2\nu} dt^2 - e^{2\psi} dr^2$. 
This gives $dt^2(1-\frac{f}{R}) = 2f,_R R^2 dr^2$. Rewriting it
we get,
\begin{equation}
 dt^2=\frac{2R^3f,_R}{(R-f)} dR^2
\end{equation}
The equation for outgoing radial null geodesics is now given by 
$dt= \pmod{[\frac{2R^3f,_R}{(R-F)}]^{\frac{1}{2}}} dR$

In the limit $R \to 0$, if the coefficient of the $dR$ 
term above is finite or zero, then the 
outgoing null geodesics from $R=0$ singularity will exist, 
and the time taken for a light ray coming out from $R=0$ to 
a very small value of $R$ is bounded. This can be seen in 
the following way. The geodesic equation in this case can be 
written as,
\begin{equation}
 dt= c(R)dR \label{ns1}
\end{equation}
If $\lim_{R \to 0}c(R)=c_0, c_0>0$, then \eqref{ns1} 
can be integrated to give $t=c_0 R+a_2$, which is the equation
of outgoing null geodesic close to the center where $a_2$ 
is some constant. If on the other hand, 
$\lim_{R \to 0}c(R)=c_1 R^n$ where $n>0$, then \eqref{ns1} 
can be integrated to give
$t=c_n \frac{R^{n+1}}{(n+1)}+a_3$. Therefore, in both 
the cases radial outgoing null geodesics coming out of the 
central singularity exist.

If in the limit $R \to 0$, either of the quantities 
$f$ or $f,_R$ has a divergence, then for a broad class of 
functions $f$, and for very small values of $R$ we can write 
O($f,_R)= $O($\frac{1}{R}$)O($f$). So, 
for small values of $R$, we have O($R-f)=$ O($f$) and 
O($R^3f,_R)=$ O($R^2$)O($f$). In that case, the 
coefficient of $dR$ is bounded because $\lim{R \to 0}$. 
This implies, as per our discussion above, that outgoing 
null geodesics do come out from the singularity.

This shows that, for a wide class of static 
massless scalar field spacetimes, the eternal singularity 
present at $R=0$, the location where the physical radius 
vanishes, is always a visible naked singularity which is 
not hidden within an event horizon.

\subsection{The $\phi=\phi(t)$ class of models}

In the comoving frame, the components of the 
energy-momentum tensor in the case when $\phi=\phi(t)$ are 
given as,
\begin{equation}
T^t_t=-T^r_r=-T^{\theta}_{\theta}=-T^{\phi}_{\phi}=
\frac{1}{2}e^{-2\ntr}\dot{\ph}^2\;\;.
\label{eq:em}
\end{equation}
Thus, we see that the massless scalar
field behaves in this frame like a {\it stiff} isentropic 
perfect fluid with the equation of state
\be
p(t,r)=\rho(t,r)=\frac{1}{2}e^{-2\ntr}\dot{\ph}^2\;\;.
\n{rhop}
\ee
In this case, we can easily see that for any real valued 
function $\ph(t)$, all energy conditions ensuring positivity
of mass and energy density of the field are satisfied by 
the energy momentum tensor.

We now have here $\phi=t$, and $g_{\mu\nu}=g_{\mu\nu}(r)$,
as we are considering the static class of solutions. 
We note that the Chase theorem does not include this case,
because for a non-empty model in this case, the scalar field 
does not go to a vanishing value faraway at large values 
of the radial coordinate, and the scalar field is not 
asymptotically flat.
To avoid any confusion, we emphasize that the condition of staticity for
the spacetime implies that the metric components are not 
functions of time. The scalar field, however, can be time-dependent 
in such a way so that the physical quantities like energy density 
and pressure are functions of $r$ only.

From the Einstein equations then it is seen that  $F=F(r)$
necessarily. 
From the Klein-Gordon equation 
we have $\dot{\phi}(t)=constant$, and we can normalise this
constant to $1$. 
Then the Einstein equations become  
\begin{equation}
\frac{1}{2}e^{-2\nu}=\frac{F'}{R^2R'} \label{t1}
\end{equation}
\begin{equation}
\frac{1}{2}e^{-2\nu} = e^{-2\psi}(\frac{R'^2}{R^2}+\frac{2R'\nu'}{R})-
\frac{1}{R^2} \label{t2}
\end{equation}
and 
\begin{equation}
e^{-2\psi} R'^2 = 1-\frac{F}{R} .
\end{equation}
Since both $e^{-2\nu}$ and $R$ are functions of $r$ only, 
therefore $e^{-2\nu}$ can always be written in the following form,
\begin{equation}
\frac{1}{2}e^{-2\nu}=\frac{f,_R}{R^2}\label{tnu}
\end{equation}
Putting this back into \eqref{t1}, we get 
\begin{equation}
F=f+C_1
\end{equation}
We take $C_1=0$ and then that gives
\begin{equation}
e^{-2\psi} R'^2 = 1-\frac{f}{R} \label{tpsi}
\end{equation}
From \eqref{tpsi}, we have
\begin{equation}
\nu'=(-\frac{f,_{RR}}{2f,_R}+\frac{1}{R})R'\label{tsub1}
\end{equation}
Using \eqref{tsub1} and \eqref{tnu} and \eqref{tpsi} in \eqref{t2}, 
we get
\begin{equation}
Rf,_R^2 = (R-f)(3f,_R-Rf,_{RR})-Rf,_R \label{tf}
\end{equation}
The solutions to the above Einstein equation
gives the classes of models for the case $\phi=\phi(t)$.

We note here that the class $\phi=\phi(t)$ for the static 
massless scalar models is non-empty, and one solution for 
this class was given by Wyman
\cite{Wyman}. 
Apart from that particular solution, it may be possible to
argue that there exist other solutions as well in this class. To consider 
this, the Einstein equations in this case can be reduced to a 
single second order ordinary differential equation \eqref{tf}. 
Given the values of $f(R)$ and $f(R),_R$ at $R=0$ as 
initial conditions, from the existence theorems, a solution 
of \eqref{tf} exists. While this requires further investigation 
which we do not go into presently, this indicates the existence 
of other solutions for this class as well.

We can now work out the Ricci scalar in this case,
which is given by,
\begin{equation}
R_c=e^{-2\nu}\dot{\phi}^2=\frac{2f,_R}{R^2}.
\end{equation} 
It is clear that unless $f,_R$ goes to zero as $R$ goes 
to zero at least as fast as $R^2$, $R_c$ will blow up at $R=0$. 
Unlike the $\phi=\phi(r)$ case, however, \eqref{tf} does not 
forbid now such a behaviour of $f$. So it is possible that 
singularity-free solutions may exist in this case
where the spacetime has no central singularity 
at $R=0$ in all cases.

Again, the equation of horizon \eqref{aph} reduces 
to  $e^{-2\psi}R'^2=0$ in this case also. Like the $\phi=\phi(r)$ 
static case, this relation gives us the apparent horizon. From 
the Einstein equations then, we have $F=R$. This implies $f=R$ 
through a procedure similar to that we followed earlier in the case 
of $\phi= \phi(r)$. Putting this in \eqref{tf}, we get 
$Rf,_R(f,_R+1)=0$. So $f(R)=R$ can be satisfied only at 
$R=0$, or if $f,_R=0$ at some positive value of $R$. 
We note here that because of energy conditions, $f,_R$ cannot 
be negative. Therefore the relation $f=R$ can be satisfied 
only at $R=0$. As in the case of $\phi= \phi(r)$, this implies 
that there is no event horizon if the spacetime is non-empty. 
We note that this is exactly the same conclusion that we reached 
earlier for the $\phi=\phi(r)$ case.


\section{Discussion}

We briefly summarize here our main results about the
spacetime singularity and the trapped surfaces in the case
of massless scalar field spacetimes.

1) For static spherically symmetric massless scalar field spacetimes, 
there is no event horizon if the spacetime is non-empty ($i.e.$ the 
density does not go to zero at any comoving coordinate radius $r$. ) 

2) For the solutions 
where the scalar field is asymptotically flat, this agrees with the result 
by Chase 
\cite{Chase}. 
For the class of solutions $\phi=\phi(t)$, the scalar field 
does not go to zero at spatial infinity, and therefore the scalar 
field is not asymptotically flat. So Chase's theorem does 
not hold for that class of solutions. However, we still 
found that event horizon does not exist for this 
class of spherically symmetric spacetimes, independently 
of the asymptotic flatness condition.

3) For the $\phi=\phi(r)$ class of scalar field models, 
there is always a curvature singularity present at $R=0$. 
It is seen that this singularity is visible in the sense that future 
directed null geodesics come out from the same. This result agrees 
with that found by Xanthopoulos and Zannias 
\cite{Xan},
as they considered spherically symmetric massless 
scalar fields which are static and asymptotically flat, in an 
arbitrary number of spacetime dimensions, and found a 
visible curvature singularity at the physical radius $R=0$.

4) We note that the issue of visibility of the 
central singularity for the solution given by Newman, Janis and 
Winicour 
\cite{JNW} 
has been considered by Virbhadra, Jhingan and Joshi 
\cite{JNWsing}. 
They found that the singularity is visible in this case.
It is to be further noted that for the class $\phi= \phi(t)$, 
there may or may not be a singularity at 
the center $R=0$, though this point requires 
some further examination.


\begin{thebibliography}{99}

\bibitem{Bergman}O. Bergmann and L. Leipnik: Phys. Rev. 107, 1157(1957)

\bibitem{Buchdal}H.A.Buchdal: Phys. Rev. 115,1325(1959)

\bibitem{Yilmaz}H.Yilmaz:Phys. Rev. 111, 1417(1958)

\bibitem{Szekeres}G.Szekeres,:Phys. Rev. 97, 212(1955)

\bibitem{Wyman}Max Wyman; Phys Rev D, Vol 24, No 4, 1981

\bibitem{Xan}B.C.Xanthopoulos and T.Zannias: Phys. Rev.D, Vol 40, No 8, 1989


\bibitem{conformal} C. G. Callan Jr., S. Coleman and R. Jackiw, 
Ann. Phys. (NY) 59, 42(1970); L. Parker, Phys. Rev. D7, 976 (1973); 
J. D. Bekenstein, Ann. phys. 82, 535 (1974); 
B. C. Xanthopoulos, J. Math. Phys. 32, 1875 (1991).


\bibitem{scal} R. Penny, Phys. Rev.174, 1578(1968);
A. I. Janis, D. C. Winicour, Phys. Rev. 186, 1729 (1969).



\bibitem{Chase} J.E.Chase: Commun.math. Phys. 19, p. 276 (1970).


\bibitem{haw}S. W. Hawking and G. F. R. Ellis, {\it The large scale
structure of space-time}, Cambridge University Press, Cambridge (1973).


\bibitem{landau} Landau \& Lifshitz, {\it Classical theory of fields}, 
p.304 (1975).


\bibitem{GJ1} P.S. Joshi, I.H. Dwivedi, Class. Quant. Grav.
{\bf 16}, p.41 (1999); R. Goswami and P. S. Joshi, Phys. Rev. {\bf D76}, 
p.084026 (2007).


\bibitem{JNW} A.I.Janis, E.T.Newman and J.Winicour; Phys.Rev.Lett. 
{\bf 20}, p.878 (1968).


\bibitem{JNWsing}K. S. Virbhadra, S. Jhingan and P. S. Joshi: 
arxiv: gr-qc/9512030v2; Int. J. Mod. Phys. {\bf D6}, p.357 (1997). 







\end{thebibliography}
\end{document}